\documentstyle[12pt,epsfig]{article}
\makeatletter
\def\slash#1{{\mathpalette\c@ncel{#1}}} 
\makeatother

\newcommand\beq{\begin{eqnarray}}
\newcommand\eeq{\end{eqnarray}}

\newcommand\la{\langle}
\newcommand\ra{\rangle}

\def\ellslash{\rlap/{\mkern-1mu \ell}}

\def\shat{\hat{s}}
\def\that{\hat{t}}
\def\uhat{\hat{u}}

\begin{document}
\vspace*{7mm}
\begin{center}
{\Large \bf 
Single Transverse-Spin Asymmetry in $pp^\uparrow \to \pi X$ and
$ep^\uparrow \to \pi X$
  \\}
\vspace{0.5cm}
 {\sc Yuji~Koike}
\\
\vspace*{0.1cm}{\it Department of Physics, Niigata University,
Ikarashi, Niigata 950--2181, Japan}
\\[0.5cm]
\parbox[t]{\textwidth}{{\bf Abstract:} 
Cross section
formulas for the single spin asymmetry in $p^\uparrow p\to \pi(\ell_T)X$ and
$e p^\uparrow \to \pi(\ell_T)X$
are derived and its characteristic features are discussed. 
}

\end{center}

\setcounter{equation}{0}

In this report we discuss the single transverse-spin asymmetry
for the pion production with large transverse momentum $A_N$ 
in $pp^\uparrow\to\pi(\ell) X$ and $p^\uparrow e \to \pi(\ell) X$
relevant for RHIC-SPIN, HERMES and COMPASS experiments. 
According to the QCD factorization theorem,
the polarized cross section for $pp^\uparrow \to \pi X$  
consists of three twist-3 contributions:
\beq
&(A)&\qquad G_a(x_1,x_2)\otimes q_b(x')\otimes D_{c}(z)\otimes
\hat{\sigma}_{ab\to c},\nonumber\\[-5pt]
&(B)&\qquad  \delta q_a(x)\otimes E_b(x_1',x_2') \otimes D_{c}(z)\otimes 
\hat{\sigma}_{ab\to c}',\nonumber\\[-5pt]
&(C)&\qquad \delta q_a(x)\otimes q_b(x') \otimes \widehat{E}_{c}(z_1,z_2)
\otimes\hat{\sigma}_{ab\to c}'',\nonumber
\eeq
where 
the functions $G_a(x_1,x_2)$, $E_b(x_1',x_2')$ 
and $\widehat{E}_{c}(z_1,z_2)$
are the twist-3 quantities representing, respectively, the
transversely polarized distribution, the unpolarized distribution, and
the fragmentation function for the pion. 
$\delta q_a(x)$ is the transversity distribution in $p^\uparrow$. 
$a$, $b$ and $c$ stand for the parton's species, sum over which
is implied.  $\delta{q}_a$, $E_b$ and $\widehat{E}_c$ are chiral-odd.  
Corresponding to the above
(A) and (C), the polarized cross section for $ep^\uparrow \to \pi X$
(final electron is not detected) receives two twist-3 contributions:
\beq
&(A')&\qquad G_a(x_1,x_2)\otimes 
D_{a}(z)\otimes\hat{\sigma}_{ea\to a}, \nonumber\\[-5pt]
&(C')&\qquad \delta q_a(x)\otimes\widehat{E}_{a}(z_1,z_2)\otimes
\hat{\sigma}_{ea\to a}'.\nonumber
\eeq
The (A) and (B) contributions for $pp^\uparrow\to \pi X$
have been analyzed in \cite{QS99} and \cite{KK00}, respectively, and
it has been shown that (A) gives rise to large $A_N$ at large $x_F$ 
as observed in E704, and (B) is negligible in all kinematic region. 
Here we extend the study to the (C) term (see also \cite{Koike01})
at RHIC energy and also the asymmetry in $ep$ collision.

The transversely polarized twist-3 distributions 
$G_{F}(x_1,x_2)$ relevant to the (A) term
is given in \cite{KK00}.
Likewise 
the twist-3 fragmentation function for the pion (with momentum $\ell$) 
is defined as the
lightcone correlation function as ($w^2=0$, $\ell\cdot w =1$)
\beq
& &{1\over N_c}\sum_X\int {d\lambda\over 2\pi}\int {d\mu\over 2\pi}
e^{-i{\lambda\over z_1}}e^{-i\mu({1\over z_2}-{1\over z_1})}
\la 0 |\psi_i(0)|\pi X \ra \la \pi
X|g F^{\alpha\beta}(\mu w)w_\beta
\bar{\psi}_j(\lambda w)|0\ra\nonumber \\
& &\qquad= {M_N\over 2z_2}\,
\left(\gamma_5\ellslash\gamma_\nu\right)_{ij}\epsilon^{\nu
\alpha\beta\lambda}w_\beta\ell_\lambda\widehat{E}_F(z_1,z_2)
+\cdots.
\label{EFhat}
\eeq
Note that we use the nucleon mass $M_N$ to normalize the twist-3
pion fragmentation function.
There is another twist-3 fragmentation function
which is obtained from (\ref{EFhat}) by shifting
the gluon-field strength from the
left to the right of the cut.
The defined function $\widehat{E}_{FR}(z_1,z_2)$ 
is connected to $\widehat{E}_{F}(z_1,z_2)$ by the relation
$\widehat{E}_F(z_1,z_2) = \widehat{E}_{FR}(z_2,z_1)$,
which follows from hermiticity and time reversal invariance.
Unlike the twist-3 distributions, the twist-3 fragmentation function
does not have definite symmetry property.

Following \cite{QS99}-\cite{Koike01}, we analyze the asymmetries
focussing on the soft-gluon pole contributions with
$G_F(x,x)$ and $\widehat{E}_F(z,z)$.
In the large $x_F$ region, i.e. production of pion 
in the forward direction of the polarized nucleon, 
the main contribution comes from the large-$x$ and large-$z$ region
of distribution and fragmentation functions, respectively. 
Since $G_F$ and $\widehat{E}_F$ behaves as $G_F(x,x)\sim (1-x)^\beta$
and $\widehat{E}_F(z,z)\sim (1-z)^{\beta'}$ with $\beta$, $\beta'>0$,
$|(d/dx)G_F(x,x)|\gg |G_F(x,x)|$,
$|(d/dz)\widehat{E}_F(z,z)| \gg |\widehat{E}_F(z,z)|$ at
large $x$ and $z$.  In particular, the valence component of
$G_F$ and $\widehat{E}_F$ dominate in this region.
We thus keep only the valence quark contribution for the derivative
of these soft-gluon pole functions (``valence-quark soft-gluon
approximation'') for the $pp$ collision.  For the $ep$ case,
we include all the soft-gluon pole contribution.

In general
$A_N$ is a function of 
$S=(P+P')^2\simeq 2P\cdot P'$,
$T=(P-\ell)^2\simeq -2P\cdot \ell$ and
$U=(P'-\ell)^2\simeq -2P'\cdot \ell$ where $P$, $P'$ and $\ell$
are the momenta of $p^\uparrow$, unpolarized $p$ (or $e$), 
and the pion respectively. 
In the following we use
$S$, $x_F = {2\ell_{\parallel}\over \sqrt{S}} = {T-U\over S}$ and
$x_T = {2\ell_{T}\over \sqrt{S}}$ as independent variables.
The polarized cross section for the (C) term reads
\beq
& &E_\pi{d^3\Delta\sigma(S_\perp) \over d \ell^3}
= {2\pi M_N \alpha_s^2 \over S}\epsilon^{\alpha\ell w S_\perp}
\sum_{a}\int_{z_{min}}^1
{d\,z\over z^2}
\int_{x_{min}}^1 {d\,x\over x}
{1\over xS + U/z}
\int_0^1 {d\,x'\over x'}
\nonumber\\
& &\qquad\times \delta\left(x'+{xT/z \over xS + U/z}\right)\nonumber\\
& &\qquad\left\{
\sum_b \delta q^a(x) q^b(x') 
\left[-z_1^2 {\partial \over \partial z_1}
\widehat{E}_{F}^a(z_1,z)\right]_{z_1=z}
\left({-2p_\alpha\over T}\widehat{\sigma}_{ab\to a}^I+
{-2p'_\alpha\over U}\widehat{\sigma}_{ab\to a}^{II}\right)\right.\nonumber\\
& &\qquad\left.+\sum_b \delta q^a(x) q^b(x') 
\left[ -z^2 {d \over d z}
\widehat{E}_{F}^a(z,z)\right]
{xp_\alpha+x'p'_\alpha \over |xT+x'U|}
\left(\widehat{\sigma}_{ab\to a}^I+
\widehat{\sigma}_{ab\to a}^{II}\right)\right.\nonumber\\
& &\qquad\left.+\delta q^a(x) G(x') 
\left[-z_1^2 {\partial \over \partial z_1}
\widehat{E}_{F}^a(z_1,z)\right]_{z_1=z}
\left({-2p_\alpha\over T}\widehat{\sigma}_{ag\to a}^I+
{-2p'_\alpha\over U}\widehat{\sigma}_{ag\to a}^{II}\right)\right.\nonumber\\
& &\qquad\left.+\delta q^a(x) G(x') 
\left[ -z^2 {d \over d z}
\widehat{E}_{F}^a(z,z)\right]
{xp_\alpha +x'p'_\alpha \over |xT+x'U|}
\left(\widehat{\sigma}_{ag\to a}^I+
\widehat{\sigma}_{ag\to a}^{II}\right)\right\},
\label{final}
\eeq
where the lower limits for the integration variables are
$z_{min} = -(T+U)/S=\sqrt{x_F^2+x_T^2}$ and 
$x_{min} = -U/z(S+T/z)$.
Using the invariants in the parton level,
$\shat =(xp + x'p')^2 =xx'S$,
$\that =(xp-{\ell/z})^2 =xT/z$ and
$\uhat =(x'p'-{\ell/z})^2 =x'U/z$,   
the partonic hard cross sections read
\beq
& &\widehat{\sigma}_{qq'\to q}^I={\shat\uhat\over 18\that^2} -
{\shat\over 54 \that}\delta_{qq'},\quad
\widehat{\sigma}_{q\bar{q'}\to q}^I={\shat\uhat\over 18\that^2},\quad
\widehat{\sigma}_{\bar{q}q'\to \bar{q}}^I=-{\shat\uhat\over 18\that^2},
\nonumber\\
& &\widehat{\sigma}_{\bar{q}\bar{q'}\to \bar{q}}^I=
-{\shat\uhat\over 18\that^2} +{\shat\over 54 \that}\delta_{qq'},
\quad
\widehat{\sigma}_{qq'\to q}^{II}=-{7\shat\uhat\over 18\that^2} -
{\shat\over 54 \that}\delta_{qq'},\quad
\widehat{\sigma}_{q\bar{q'}\to q}^{II}
=-{\shat\uhat\over 9\that^2},\nonumber\\
& &\widehat{\sigma}_{\bar{q}q'\to \bar{q}}^{II}
={\shat\uhat\over 9\that^2},\qquad
\widehat{\sigma}_{\bar{q}\bar{q'}\to \bar{q}}^{II}
={7\shat\uhat\over 18\that^2} +{\shat\over 54 \that}\delta_{qq'},\nonumber\\
& &
\widehat{\sigma}_{qg\to q}^{I}={\shat\uhat\over 8\that^2} 
+{\shat+\uhat\over 16\that} +{5\over 72},\qquad
\widehat{\sigma}_{qg\to q}^{II}=-{9\shat\uhat\over 16\that^2} 
-{9\uhat\over 16\that} -{1\over 16}.
\label{parton}
\eeq
At large $x_F$ ($-U \gg -T$), 
$\sigma^I$ becomes more important because of $1/T$ factor
in (\ref{final}).

\begin{figure}[htb]
\setlength{\unitlength}{1cm}
\begin{picture}(14,5)(0,-0.25)
\psfig{file=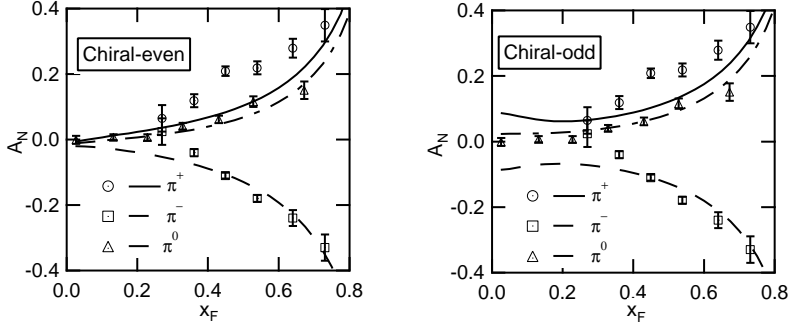,height=5cm}
\end{picture}\par
\vspace*{-0.85cm}
\caption{$A_N^{pp}$ at $\sqrt{S}=20$ GeV and $\ell_T=1.5$ GeV.}
\end{figure}

To estimate the above contribution,
we introduce a simple model ansatz as
$\widehat{E}_F^a(z,z)=K_a D_{a}(z)$ 
with a flavor dependent factor $K_a$.
$K_a$'s are determined to be $K_u=-0.11$ and $K_d=-0.19$ 
so that (\ref{final}) approximately gives rise to $A_N^{pp}$ observed
in E704 data at $\sqrt{S}=20$ GeV and $\ell_T=1.5$ GeV.  As noted before, 
$\widehat{E}_F(z_1,z_2)$ does not have definite symmetry property
unlike the twist-3 distribution $G_F(x_1,x_2)$.
Nevertheless we assume 
$\left[(\partial/\partial z_1)\widehat{E}_F(z_1,z)\right]_{z_1=z}
=(1/2)(d/d z)\widehat{E}_F(z,z)$.  We refer the readers to \cite{KK00}
for the adopted distribution and fragmentation functions.
The result for $A_N^{pp}$ from the (C) (chiral-odd) term
is shown in 
Fig. 1 in comparison with the (A) (chiral-even) contribution.
(See \cite{KK00} for the detail.)
Both effects give rise to $A_N^{pp}$ similar to the E704 data.
The origin of the growing
$A_N$ at large $x_F$ is
(i) large partonic cross sections in (\ref{parton}) ($\sim 1/\hat{t}^2$ term)
and (ii) the derivative of the soft-gluon pole functions.  
With the parameters $K_a$ fixed,
$A_N^{pp}$ at RHIC energy ($\sqrt{S}=200$ GeV)
is shown in Fig.2 at $l_T=1.5$ GeV.  
Both (A) and (C) contributions give slightly 
smaller $A_N^{pp}$ than the STAR data
reported at this conference\,\cite{STAR}.  
Fig.3 shows the $A_N^{pp}$ as a function of $\ell_T$ at $\sqrt{S}=200$ GeV and
$x_F=0.6$, indicating quite
large $l_T$ dependence in both (A) and (C) contributions
at $1< l_T < 4$ GeV region.

\begin{figure}[htb]
\setlength{\unitlength}{1cm}
\begin{picture}(14,5)(0,-0.2)
\psfig{file=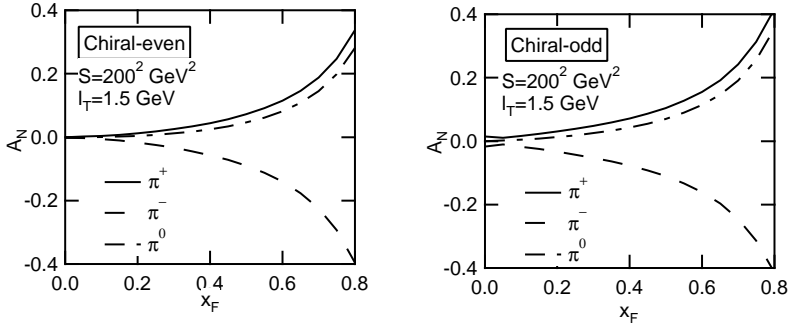,height=5cm}
\end{picture}\par
\vspace*{-0.65cm}
\caption{$A_N^{pp}$ at $\sqrt{S}=200$ GeV and $\ell_T=1.5$ GeV.}
\end{figure}

We next discuss the asymmetry $A_N^{ep}$ for $p^\uparrow e\to \pi(\ell)X$
where the final electron is not observed.  
In our $O(\alpha_s^0)$ calculation, 
the exchanged photon remains highly virtual
as far as the observed $\pi$ has
a large trasverse momentum $\ell_T$ with respect to the $ep$ axis.
Therefore experimentally one needs to integrates only over those 
virtual photon events to compare with our formula.

\begin{figure}[hbtp]
\setlength{\unitlength}{1cm}
\begin{minipage}[t]{7cm}
\begin{picture}(7,5)(0,-0.2)
\psfig{file=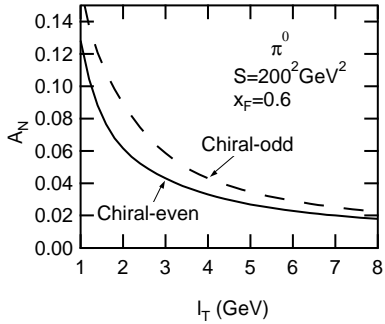,height=5cm}
\end{picture}\par
\end{minipage}
\hfill
\begin{minipage}[t]{7cm}
\hspace*{-1cm}
\vspace*{-4cm}
\caption{$\ell_T$ dependence of $A_N^{pp}$ at $\sqrt{S}=200$ GeV and 
$x_F=0.6$.}
\end{minipage}
\end{figure}

\vspace{-0.75cm}

Using the twist-3 distribution and fragmentation functions 
used to describe $pp$ data, we show in Fig. 4 
$A_N^{ep}$ corresponding to
(A')(chiral-even) and (C')(chiral-odd) contributions.
Remarkable feature of Fig. 4 is that in both chiral-even and
chiral-odd contributions (i) the sign of $A_N^{ep}$
is opposite to the sign of $A_N^{pp}$ and (ii) the magnitude of $A_N^{ep}$
is much larger than that of $A_N^{pp}$, in particular, at large $x_F$,
and it even overshoots one.
(In our convention, $x_F >0$ 
corresponds to the production of $\pi$ in the forward hemisphere of
the initial polarized proton both in $p^\uparrow p$ and $p^\uparrow e$ case.)
The origin of these features can be traced back to the color factor in
the dominant diagrams for the {\it twist-3 polarized} cross sections
in $ep$ and $pp$ collisions.
Of course, the asymmetry can not exceeds one, and thus our model 
estimate needs to be modified.  
First, the applied kinematic range
of our formula should be reconsidered: Application of the twist-3 cross section
at such small $\ell_T$ may not be justified.
Second, our model ansatz of $G_F^a(x,x)\sim q^a(x)$ and 
$\widehat{E}_F^a(z,z)\sim D^a(z)$ is not appropriate at $x\to 1$ and
$z\to 1$, respectively.  The derivative of these functions,
which is important for the growing $A_N^{pp}$ at large $x_F$, 
eventually
leads to divergence of $A_N^{pp}$ at $x_F\to 1$ as $\sim 1/(1-x_F)$.

\begin{figure}[htb]
\setlength{\unitlength}{1cm}
\begin{picture}(14,5)(0,-0.2)
\psfig{file=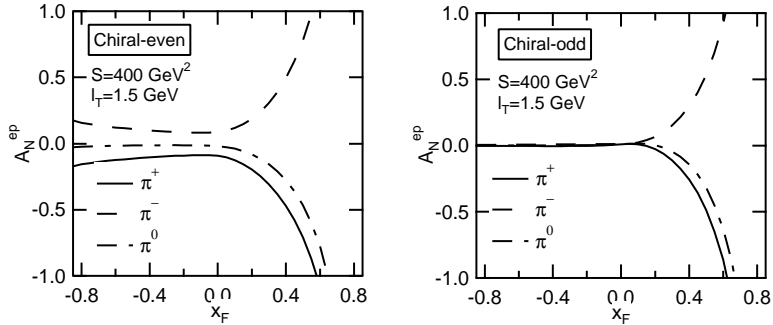,height=5cm}
\end{picture}\par
\vspace*{-0.8cm}
\caption{$A_N^{ep}$ at $\sqrt{S}=20$ GeV.}
\end{figure}

As a possible remedy to
this pathology we tried the following.
For the (A) (chiral-even)
contribution
we have a model $G_F^a(x,x)\sim q_a(x)\sim_{x\to 1} (1-x)^{\beta_a}$
where $\beta_u=3.027$ and $\beta_d=3.774$ in the GRV distribution we adopted.
Tentatively we shifted $\beta_{u,d}$ as $\beta_a\to\beta_a(x)=\beta_a+x^3$,
which suppresses the divergence of $A_N$ at $x_F\to 1$ but still
causes rising behavior of $A_N$ at large $x_F$.  This avoids 
overshooting of one in $A_N^{ep}$ but reduces $A_N^{pp}$, typically 
by factor 2.  So the twist-3 contribution to $A_N^{pp}$
shown in Fig. 1 and 2 is further reduced, making the deviation from E704 and
STAR data bigger.  

To summarize we have studied the $A_N$ for pion
production in $pp$ and $ep$ collisions.  Although our approach
provides a systematic framework for the large $\ell_T$ production,
applicability of the formula to the currently available low $\ell_T$
data still needs to be tested, in particular, 
comparison of $\ell_T$ dependence with data is needed.

\vspace{0.2cm}
\noindent
{\bf Acknowledgement:}  I would like to thank Daniel Boer, Jianwei Qiu and
George Sterman for useful discussions.  This work is supported in part 
by the Grant-in-Aid for Scientific Research of Monbusho.


\vspace{-0.5cm}

\small

\begin{thebibliography}{99}


\bibitem{QS99} J. Qiu and G. Sterman, Phys. Rev. {\bf D59} (1999) 014004.

\bibitem{KK00} Y. Kanazawa and Y. Koike, Phys. Lett. {\bf B478} (2000) 121;
{\bf B490} (2000) 99.

\bibitem{Koike01} Y. Koike, hep-ph/0106260 (Proceedings of DIS2001, Bologna, 
Italy, April, 2001.)

\bibitem{STAR} G. Rakness, these proceedings.
\end{thebibliography}
\end{document}